# Data Science: A Three Ring Circus or a Big Tent?

Jennifer Bryan[1] and Hadley Wickham[2]

*Note: This is part of a collection of discussion pieces on David Donoho's paper* 50 Years of Data Science.

- *Main article "50 Years of Data Science":*
  *https://doi.org/10.1080/10618600.2017.1384734*
- *Discussion pieces are in Volume 26, Issue 4 of the Journal of Computational and Graphical Statistics (2017):* http://www.tandfonline.com/toc/ucgs20/26/4

We enjoyed re-reading this important article from Donoho and are pleased to see it discussed here in JCGS.

For context, we both trained as statisticians and spent several years as regular professors of Statistics. We still have academic appointments. Yet today we work for RStudio, building tools to improve the workflows for data scientists and statisticians. This gives us an informed and unique perspective on Donoho's piece, which explores aspects of the academic statistical establishment that are deeply connected with this unusual career path.

Overall, much of the paper resonated with us. Our comments deal with three main areas: academic statistics, the skills meme, and coupling of cognitive and computation tools.

## Academic statistics

Donoho gives a beautiful synthesis of the (largely-unheeded) pleas from four distinguished statisticians, who, over 50 years, argued for an expanded definition of "academic statistics". He rightly points out that statisticians and departments of Statistics generally do not lead the Data Science initiatives at major universities. But Donoho stops short of making the obvious connection: maybe there is a causal relationship between the two facts? Perhaps the reluctance to embrace data preparation, presentation, and prediction is precisely why Statistics often finds itself on the periphery. If Statistics is unwilling to own the full range of activities necessary to learn from data, how is it possible to claim that "Data Science is just statistics"?

Anyone who has ever taken wild-caught data through the full process of analysis knows that "statistics", in the strict sense of fitting models and doing inference, is but one small part of the process. Every repetition of the sentiment that "Data Science is just statistics" underscores how many statisticians have yet to appreciate and accept the changes going on around them. It is understandable that Statistics departments want to share in the resources flowing to Data Science Initiatives, but that comes with the responsibility to address a broader set of needs.

---

[1] RStudio, University of British Columbia

[2] RStudio, Stanford University, University of Auckland, Rice University

This unnecessarily narrow definition of our field is often paired with a narrow definition of who is allowed to do statistics. Statisticians have a tendency to advocate statistical abstinence: you should only practice statistics if you're in a long-term relationship with a statistician (Wickham 2015). But abstinence-based education rarely works. People see their friends using statistics and having a great time, and there simply aren't enough statisticians to go around. As a field, we need to teach safe-stats, not just statistical abstinence.

Donoho correctly confirms that applied statisticians regularly engage in all the activities touted in press releases, like the one he quotes: "the collection, management, processing, analysis, visualization, and interpretation of vast amounts of heterogeneous data associated with a diverse array of … applications." But there is currently a big gap between what statisticians *do* and what is considered worthy of *study*. The incentive structures of academic statistics still signal that mathematical statistics and the creation of new models and inferential procedures are more valuable than work related to data manipulation, visualisation, and programming. This is reflected in the content of for-credit courses, qualifying exams, and standards for funding and promotion. Graduate students and junior faculty are caught between a rock and a hard place (Waller 2017). It can be very difficult to present modern data scientific work as impactful scholarly activity, when the system still defines that primarily as theory and methodology papers.

The good news is the above dysfunction is largely self-imposed! The constraints on what Statistics means are coming from inside the house. Donoho's Six Divisions of Greater Data Science make a wonderful basis for evaluating the usefulness of various activities. We hope that academic departments, grant panels, and promotion committees start to use them.

## The skills meme and GDS3: Computing with Data

In "The 'Skills' Meme", Donoho sets out to debunk claims about specific skills that allegedly distinguish Data Science. We'd like to refine the points made about Hadoop, "a variant of Map/Reduce for use with datasets distributed across a cluster of computers". We think Hadoop is mostly transitional: it is important for *someone* to worry about the issues of data storage and managing computation, in the same way *someone* should worry about measure theory. Unless you're dealing with exceptionally large data (i.e. multiple terabytes), data representation isn't something that most data scientists need to think about: worrying about data storage and sharded computation is becoming the responsibility of a cadre of specialised data engineers.

We want to underscore that there are substantive skills that distinguish Data Science training from that currently offered in Statistics. This relates to what Donoho refers to as GDS3: Computing with Data. Yes, Data Science tends to place a greater emphasis on computing and programming. But statisticians are too quick to discount this as fairly superficial issues of mechanics, like re-writing R code to avoid using for-loops. The skills meme actually runs much deeper when it comes to computation and programming. It is related to important issues of correctness and reusability. There are proven frameworks from software engineering and IT operations that need to become much more common in data analysis (Parker 2017). We must acknowledge the kernel of truth in the cringe-worthy term "professor-ware".

We don't claim that all statisticians should write code that is ready for others to use. But surely some should! The basic practices of modularity, testing, version control, packaging, and interface design are not mere niceties. They determine whether data scientific products can actually be trusted and built upon, like a proof in mathematics. It is accepted that we should exploit the methodological innovations made in the past 15 years. Likewise, we must acknowledge big changes in the standards for modern scientific computing. If the era of Data Science prompts a long-overdue enlargement of Statistics, we would do well to incorporate these valuable skills into our revamped curriculum.

## Coupling cognitive and computational tools

Mathematics provides a common language for mathematical statistics. For exactly the same reasons, it is vital to have shared abstractions and notation when solving problems in applied statistics. This is what a programming language provides, i.e. it is not just syntax for issuing instructions to a computer. Although a programming language cannot be as timeless as mathematics, R currently provides a powerful language for applied statistics.

Donoho generously mentions the benefits of the R packages reshape2 and plyr. These are early milestones in an effort that has more recently matured into the so-called Tidyverse, https://www.tidyverse.org, an ecosystem of packages designed for data science. In ggplot2 and dplyr, the Tidyverse provides two illustrations of the idea that programming is a valid medium for intellectual work and human communication. ggplot2 and dplyr are clear intellectual contributions because they provide tools (grammars) for visualisation and data manipulation, respectively. The tools make the tasks radically easier by providing clear organising principles coupled with effective code. Users often remark on the ease of manipulating data with dplyr and it is natural to wonder if perhaps the task itself is trivial. We claim it is not. Many probability challenges become dramatically easier, once you strike upon the "right" notation. In both cases, what feels like a matter of notation or syntax is really a more about exploiting the "right" abstraction.

Another part of what makes the Tidyverse effective is harder to see and, indeed, the goal is for it to become invisible: conventions. The Tidyverse philosophy is to rigorously (and ruthlessly) identify and obey common conventions. This applies to the objects passed from one function to another and to the user interface each function presents. Taken in isolation, each instance of this seems small and unimportant. But collectively, it creates a cohesive system: having learned one component you are more likely to be able to guess how another different component works.

The Tidyverse explicitly recognizes that technology, especially programming, is part of the problem domain. It doesn't matter how good a theoretical solution is, unless there are practical tools that implement it. We must also recognise that humans are an essential part of the data science process and study how they can interact with the computer most effectively. Finding useful abstractions and exposing them through programming languages is an important part of this process.

## Conclusion

We appreciate this opportunity to comment on the important issues Donoho has raised for the next 50 years of Statistics. Readers can keep exploring these topics in Practical Data Science for Stats https://peerj.com/collections/50-practicaldatascistats/, a collection of articles we've co-edited as a PeerJ preprint Collection and a future special issue of The American Statistician.

We see a substantial mismatch between what is needed to learn from data and the much smaller subset of activity that is structurally rewarded in academic statistics today. We both still love to teach and to let those experiences inform the design of better tools and workflows for data analysis. But, frankly, this currently feels easier to do outside the academy.

Data Science has at least one advantage over Statistics, which partially explains its existence. Re-defining an existing field like Statistics is terribly difficult, whereas it's much easier to define something new from scratch. Increasing activity in the areas proposed by Donoho inevitably means reducing the traditional supremacy of statistical theory. It remains to be seen whether the community has the will to finally heed the call of Tukey, Chambers, Cleveland and Breiman, and rethink our priorities.

## References

Parker, Hilary. 2017. "Opinionated Analysis Development." *PeerJ Preprints* 5 (August): e3210v1. doi:10.7287/peerj.preprints.3210v1.

Waller, Lance A. 2017. "Documenting and Evaluating Data Science Contributions in Academic Promotion in Departments of Statistics and Biostatistics." *PeerJ Preprints* 5 (August): e3204v1. doi:10.7287/peerj.preprints.3204v1.

Wickham, Hadley. 2015. "Teaching Safe-Stats, Not Statistical Abstinence." *Online Supplement Discussion of "Mere Renovation Is Too Little Too Late: We Need to Rethink Our Undergraduate Curriculum from the Ground Up" by G. Cobb, the American Statistician* 69. http://nhorton.people.amherst.edu/mererenovation/17_Wickham.PDF.